\documentclass[%
 reprint,
superscriptaddress,
]{revtex4-2}

\usepackage{tikz}
\usetikzlibrary{calc}
\usetikzlibrary{shapes.geometric, arrows}
\usepackage{eufrak}
\usepackage{enumitem}
\usepackage{mathtools}
\usepackage{graphicx}
\usetikzlibrary{positioning}
\usepackage{bm}
\usepackage{csquotes}
\usepackage{tabularx}
\newcommand{\bematrix}{\left(\begin{matrix}}
\newcommand{\ematrix}{\end{matrix}\right)}
\usepackage{ulem} % only here for strikethrough-font
\usepackage[caption=false]{subfig}
\usepackage{dcolumn}
\usepackage{amsmath,amssymb}
\usepackage{bm}
\usepackage{bbm}
\usepackage{overpic}
\usepackage{latexsym}
\usepackage{color}
\usepackage[english]{babel}
\usepackage{latexsym}
\usepackage{psfrag,graphicx}
\usepackage{epsf}
\usepackage{amsmath}
\usepackage{amssymb}
\usepackage{amsfonts}
\usepackage{natbib}

\usepackage{appendix}
\usepackage{verbatim}
\usepackage{enumitem}
\usepackage{dsfont}
\usepackage{float}
\usepackage{tikz}

\definecolor{mygrey}{gray}{0.35}
\definecolor{myblue}{rgb}{0.2,0.2,0.8}
\definecolor{myzard}{cmyk}{0,0,0.05,0}
\definecolor{mywhite}{rgb}{1,1,1}
\definecolor{myred}{rgb}{0.9,0.1,0.}
\usepackage[colorlinks=true,citecolor=myblue,linkcolor=myblue,urlcolor=myblue]{hyperref}

\usepackage[makeroom]{cancel}

\newenvironment{proof-of}[1]{\medskip\noindent\textbf{Proof of {#1}.}}{\hfill$\blacksquare$\medskip}
\newcommand{\ket}[1]{\left\vert#1\right\rangle}

\usepackage{tcolorbox}

\begin{document}

\preprint{APS/123-QED}

\title {Emergent Bell Phase in an Electro-Nanomechanical Quantum Simulator}

\author{David Ullrich}
\affiliation{Física Teòrica: Informació i Fenòmens Quàntics, Departament de Física, Universitat Autònoma de Barcelona, 08193 Bellaterra, Spain}
\author{Marta Cagetti}
\affiliation{ICFO - Institut De Ciencies Fotoniques, The Barcelona Institute of Science and Technology, 08860 Castelldefels, Barcelona, Spain}
\author{Stefan Forstner}
\affiliation{ICFO - Institut De Ciencies Fotoniques, The Barcelona Institute of Science and Technology, 08860 Castelldefels, Barcelona, Spain}
\author{Adrian Bachtold}
\affiliation{ICFO - Institut De Ciencies Fotoniques, The Barcelona Institute of Science and Technology, 08860 Castelldefels, Barcelona, Spain}
\author{Anna Sanpera}
\affiliation{Física Teòrica: Informació i Fenòmens Quàntics, Departament de Física, Universitat Autònoma de Barcelona, 08193 Bellaterra, Spain}
\affiliation{ICREA, Pg. Lluís Companys 23, 08010 Barcelona, Spain}

\date{\today}% It is always \today, today,

\begin{abstract}

Suspended carbon nanotubes hosting electrostatically defined quantum dots allow for exceptionally strong and tunable electromechanical coupling as well as mechanical modes that can reach the quantum ground state of motion simply by cryogenic cooling. This makes them a unique platform for quantum simulation of electron-phonon coupling. Here, we propose an experimentally realizable setup with two such carbon nanotubes in parallel, each hosting four quantum dots. Our system not only exhibits phonon-mediated electron–electron attraction, but also supports a robust, maximally entangled Bell phase at mesoscopic scales shared across the subsystems. These features highlight its potential as a simulator of strongly correlated quantum systems.

\end{abstract}

\maketitle

%\tableofcontents

\noindent\textit{Introduction --} Quantum electro-nanomechanical systems are increasingly employed as a toolbox for engineering and analyzing quantum phenomena at mesoscopic scales \cite{Bachtold_Meso_2022,guo_optical_2021,PRXQuantum.4.020329}. Such devices open a window for tailoring mesoscopic superpositions \cite{Romero-Isart_Superposition_2011,Potts_Phonon_2025}, for tracking gravitational forces at minute scales \cite{Gely_Diosi_2021,Omahen_Mechanical_2025}, or for simulating and probing strongly correlated physics \cite{poot2012mechanical,lewenstein_ultracold_2012} at precisions that can hardly be achieved in other platforms.

Suspended carbon nanotubes (CNTs) hosting quantum dots (QDs) represent the smallest and lightest solid-state electromechanical platforms developed to date. Their exceptionally low mass leads to large mechanical zero-point motion, making them highly promising candidates for the exploration of mechanical quantum phenomena at the mesoscale. In a CNT, QDs are created and controlled by placing gate electrodes beneath the nanotube, enabling the confinement of electrons within potential wells that exhibit discrete energy levels \cite{tormo-queralt_novel_2022, vigneau_ultrastrong_2022}. 
Single-electron tunneling into one suspended QD results in large backaction on the nanotube vibrations~\cite{steele2009strong,lassagne2009coupling,benyamini2014real},
establishing suspended CNTs as an excellent platform for investigating electron-phonon interactions \cite{bhattacharya_phonon-induced_2021, zhang_steady-state_2023}. Another key advantage of CNT-based QD systems is their ability to reach the ultra-strong coupling regime, in which the coupling between the mechanical modes of the nanotube and electron tunneling through the QD exceeds the mechanical frequency \cite{vigneau_ultrastrong_2022, Samanta_Nonlinear_2023}. Altogether, these features position CNTs as ideal candidates for simulating \textit{strongly} correlated electron-phonon (SCEP) systems.

 SCEP models are crucial in the study of quantum materials, where the interplay between mobile electrons and phonons (i.e., quantized lattice vibrations), along with electron-electron Coulomb interactions, is expected to govern both mechanical and electronic properties of the systems. Such a complex interaction landscape is believed to underlie the emergence of exotic quantum phases present in unconventional superconductivity in aromatic superconductors \cite{mitsuhashi2010superconductivity}  and alkali-metal-doped fullerides \cite{takabayashi2009disorder} as well as exotic electronic behavior in twisted bilayer graphene \cite{andrei_graphene_2020}. Note that such a simulation cannot be achieved in other platforms, like e.g., generic ultracold atomic platforms, where phonons are absent due to the rigidity of optical lattices \cite{lewenstein_ultracold_2007}.

\begin{figure}
    \vspace{-0.5cm}
    \centering
    \includegraphics[width=\linewidth]{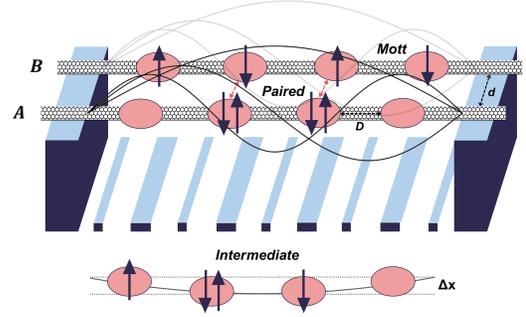}
    \vspace{-1.3cm}
    \caption{\textit{Sketch of the proposed quasi 2-D setup}. Each CNT, labeled $\mathcal{A}$ and 
    $\mathcal{B}$, hosts four quantum dots at half filling. The electronic states are capacitively coupled to the vibrational modes of the carbon nanotubes via gate electrodes located at the bottom of the trench. This results in configuration-dependent displacements ($\Delta x$). Coulomb interactions (indicated by pink arrows) occur between opposing occupied QDs on different CNTs. We assume $D > d$, i.e., the distance between neighbouring QDs on a single tube is greater than the distance between the tubes.
    Depending on the Hamiltonian parameters, various electronic configurations can emerge, including Mott insulating,  Paired and Intermediate states.}
    \vspace{-0.1cm}
    \label{fig:schematic}
\end{figure}

From a theoretical perspective, electron-phonon models are extremely challenging to tackle. On the one hand, understanding the macroscopic properties of quantum materials requires analyzing two- or three-dimensional many-body systems, significantly increasing computational complexity. On the other hand, reaching the strong electron-phonon coupling regime entails dealing with an unbounded Hilbert space due to the infinite number of phonon excitations. Altogether, that makes the simulation on a classical computer unfeasible, and quantum-inspired algorithms have been proposed to address them for very small system sizes \cite{Denner2003}.

Previous theoretical studies of an array of four QDs in a single-CNT system have demonstrated the stabilisation of electronic configurations due to phonon-mediated attractive electron-electron interactions \cite{bhattacharya_phonon-induced_2021, zhang_steady-state_2023}. In contrast, we focus on phenomena that emerge only when two parallel CNTs are present, going beyond the one-dimensional setup. This enables the generation of maximally entangled quantum states in mesoscopic systems, providing a significant step towards the experimental realization of quantum simulators for quantum materials. In what follows, we first introduce our model, consisting of two CNTs, each populated by four QDs and described by a Hubbard–Fröhlich-type Hamiltonian. We then analyze the system analytically in the zero-tunneling limit using the Lang–Firsov approximation. For finite tunneling, however, the ground state of the system must be obtained through numerical diagonalization of the Hamiltonian. As shown below, our results reveal the presence of a robust Bell phase.\\
%This paper is divided into three main sections. We begin by presenting the theoretical model used to describe the system, including its Hamiltonian. The results are then divided into two parts: analytical results obtained in the zero-tunneling limit and numerical results for the finite-tunneling regime. Finally, to provide a more detailed explanation of the numerical methods employed, we include a Supplementary Material section.}\\

\noindent\textit{The model --} We consider two identical suspended CNTs in parallel, each containing four equally spaced QDs, occupied by four unpolarized electrons (half-filling),
as indicated in Fig. \ref{fig:schematic}. 
On each individual tube, electrons can tunnel between neighbouring QDs, %at a rate $t$, 
experience on-site Coulomb repulsion 
and interact with the vibrational modes (i.e., phonons) of the suspended CNT. 
We further assume that on each CNT, inter-site Coulomb repulsion is negligible compared to on-site Coulomb repulsion.

If, as depicted in Fig. \ref{fig:schematic}, the separation between the CNTs $d$ is smaller than the inter-tube QD-separation $D$, then the dominant inter-tube coupling arises from Coulomb interactions between opposing QDs. 
%(see Fig.\ref{fig:schematic}).
The full Hamiltonian of the system has the form:  

\begin{align}
    \hat H^{\scriptscriptstyle \mathcal{AB}} = \hat H^{\scriptscriptstyle \mathcal{A}} + \hat H^{\scriptscriptstyle \mathcal{B}} + \underbrace{V\sum_{i}\hat n_i^{\scriptscriptstyle \mathcal{A}}\hat n_i^{\scriptscriptstyle \mathcal{B}}}_{\hat H_{V}}
    \label{fullHamil}
\end{align}
%\vspace{-0.35cm}

where each tube Hamiltonian reads:

\begin{align}
    \hat H^{\scriptscriptstyle \mathcal{A}/\mathcal{B}} &= -t\sum_{i,\sigma}(\hat c_{i,\sigma}^{\dagger}\hat c_{i+1,\sigma} + h.c.)
   +\sum_{\mu} {\omega_{\mu}}{\hat a_{\mu}^{\dagger}}{\hat a_{\mu}} \\ \nonumber
   &+\underbrace{U\sum_{i,\sigma\neq\sigma'}\hat n_{i,\sigma}\hat n_{i,\sigma'}}_{\hat H_{U}}+ 
   \underbrace{\sum_{i,\mu}{g_{i,\mu}}{\hat n_{i}} ({\hat a_{\mu}}^{\dagger} + \hat a_{\mu})}_{\hat H_{e-p}}
   \label{SH}
\end{align}
%\vspace{-0.35cm}

\noindent The first term accounts for electron hopping at rate $t$ between neighbouring QDs, with electron creation and annihilation operators,  $\hat c_{i,\sigma}^{\dagger}$, $\hat c_{i,\sigma}$. Here, $i$ labels the QD's position and $\sigma$ the polarisation of its hosted electrons (i.e., up or down).
The term labeled $\hat H_U$ is the on-site Coulomb repulsion with $n_{i, \sigma}=\hat c_{i,\sigma}^{\dagger}\hat c_{i,\sigma}$ whose strength $U$ is assumed to be the same on each QD. The polarisation degree of freedom is only relevant in this term, since the Pauli exclusion principle forbids double occupancy in the same QD if the electrons are spin-polarised. 
The remaining two terms in $\hat H^{\scriptscriptstyle \mathcal{A}/\mathcal{B}}$, incorporate the phononic (bosonic) modes with creation and annihilation operators $\hat a^{\dagger}_{\mu}, \hat a_{\mu}$. These modes correspond to the \enquote{flexural} modes of the CNTs, i.e., an infinite sum of harmonic oscillators with number density $\hat n_{\mu}=\hat a^{\dagger}_{\mu} \hat a_{\mu}$ and frequency $\omega_{\mu}$. The term labeled $\hat H_{e-p}$ describes the electron-phonon coupling. We consider the experimentally relevant guitar-string limit for the description of the vibrations, meaning that the modes are roughly integer multiples $\mu$ of the fundamental mode $\omega_0$ ($\omega_\mu\approx \mu\,\omega_0$). Electromechanical coupling arises from the modulation of the electrostatic potential landscape by the CNT’s displacement, as the position-dependent gate capacitance shifts the energy levels of the QD. The equation describing this coupling is $g_{i,\mu} = g_0 \frac{8}{\pi}  \mu ^{-3/2} \sin\left[\pi \mu (2i - 1)/8 \right] \sin\left[\pi \mu/8 \right]$ where $g_0$ is the tunable coupling constant \cite{micchi_mechanical_2015}. We further assume that the number of electrons of each CNT is fixed (see \cite{zhang_steady-state_2023} for the effect of inclusion of a chemical potential). Note that the single CNT Hamiltonian includes all the components of SCEP systems: a Hubbard model together with the electron-phonon coupling of the Fröhlich-type \cite{alexandrov_frohlich-coulomb_2002}. The full Hamiltonian is completed by adding the term $\hat H_V$, which describes an inter-tube Coulomb repulsion between opposing QDs with interaction strength $V$, ensuring a quasi two-dimensional character in the system. \\

The Hamiltonian of our platform offers a remarkable degree of tunability through experimentally accessible parameters. In practice, the fundamental frequency $\omega_0/2\pi$ can be tuned from $10~\mathrm{MHz}$ to $1~\mathrm{GHz}$ by varying the tube length during fabrication. The on-site potential $U/(2\pi\hbar)$ typically lies in the range of $2$--$20~\mathrm{THz}$, depending on the size of the QD, which is determined by the dimensions of the underlying gate electrodes. The intertube Coulomb potential $V/(2\pi\hbar)$, which is mainly set by the intertube spacing, can reach values as large as $\sim 10^3~\mathrm{GHz}$. The coupling constant $g_0/(2\pi\hbar)$, which ranges from $0.01$ to $1~\mathrm{GHz}$, can be tuned either at the fabrication stage, by changing the distance between the CNT and the gate electrodes or by modifying the size of the gate electrode beneath the dot, or \textit{in situ}, by adjusting the voltage applied to this gate electrode. Finally, the electronic tunneling amplitude $t/(2\pi\hbar)$, typically between $1$ and $100~\mathrm{GHz}$, can be adjusted by varying the voltage applied to the gate electrodes between the dots, thereby modifying the electrostatic potential that controls the interdot coupling.\\

%In practice, $\omega_0/2\pi$ is adjustable from  $10~\mathrm{MHz}$ to $1~\mathrm{GHz}$; $t/(2\pi\hbar)$ can be set anywhere between $1$ and $100~\mathrm{GHz}$; $g_0/(2\pi\hbar)$ can be engineered in the $0.01$–$1~\mathrm{GHz}$ range;  $U/(2\pi\hbar)$ typically falls between $2$ and $20~\mathrm{THz}$, and $V/(2\pi\hbar)$—largely determined by the intertube spacing—can reach values up to $\sim 10^3~\mathrm{GHz}$. These parameters can be tuned both at the fabrication stage and \textit{in situ} during the measurement via gate-voltage control. In particular, $t/(2\pi\hbar)$ can be adjusted by varying the voltage applied to the gate electrodes between the dots, thereby modifying the electrostatic potential that controls the interdot coupling.

%The Hamiltonian of our platform provides a remarkable degree of tunability through experimental knobs. In practice, $\omega_0/2\pi$ is adjustable from  $10~\mathrm{MHz}$ to $1~\mathrm{GHz}$; $t/(2\pi\hbar)$ can be set anywhere between $1$ and $100~\mathrm{GHz}$; $g_0/(2\pi\hbar)$ can be engineered in the $0.01$–$1~\mathrm{GHz}$ range;  $U/(2\pi\hbar)$ typically falls between $2$ and $20~\mathrm{THz}$, and $V/(2\pi\hbar)$—largely determined by the intertube spacing—can reach values up to $\sim 10^3~\mathrm{GHz}$. These parameters are tunable both at the fabrication stage and \textit{in situ} during measurement via gate-voltage control.
A quantum simulator based on two nearby parallel CNTs is within near-term experimental reach. Recent advances in deterministic CNT stamping allow placement at predefined locations with spatial precision better than $100~\mathrm{nm}$~\cite{butzerin_design_2024}. A four-quantum-dot configuration per CNT is compatible with standard electrostatic gating on extended nanotubes, building on recent multi-QD devices with up to three QDs in series~\cite{tormo-queralt_novel_2022}.\\

\noindent \textit{Methods and Results --} Given the size of the Hilbert space associated with the Hamiltonian, direct diagonalisation is not possible. Analytical insight into the physics
of the problem can be obtained using the (unitary) Lang-Firsov (LF) approximation, $\hat S= \sum_{i, \mu}\frac{g_{i,\mu} \hat n_{i}}{\omega_{\mu}} ({\hat a_{\mu}}^{\dagger} - \hat a_{\mu}) $,  widely used in condensed matter physics to describe polaron physics,  which provides an exact description of the system in the atomic limit \cite{lang_kinetic_1963, bhattacharya_phonon-induced_2021}:

\begin{figure*}[t!]
    \centering
    \includegraphics[width=0.9\linewidth]{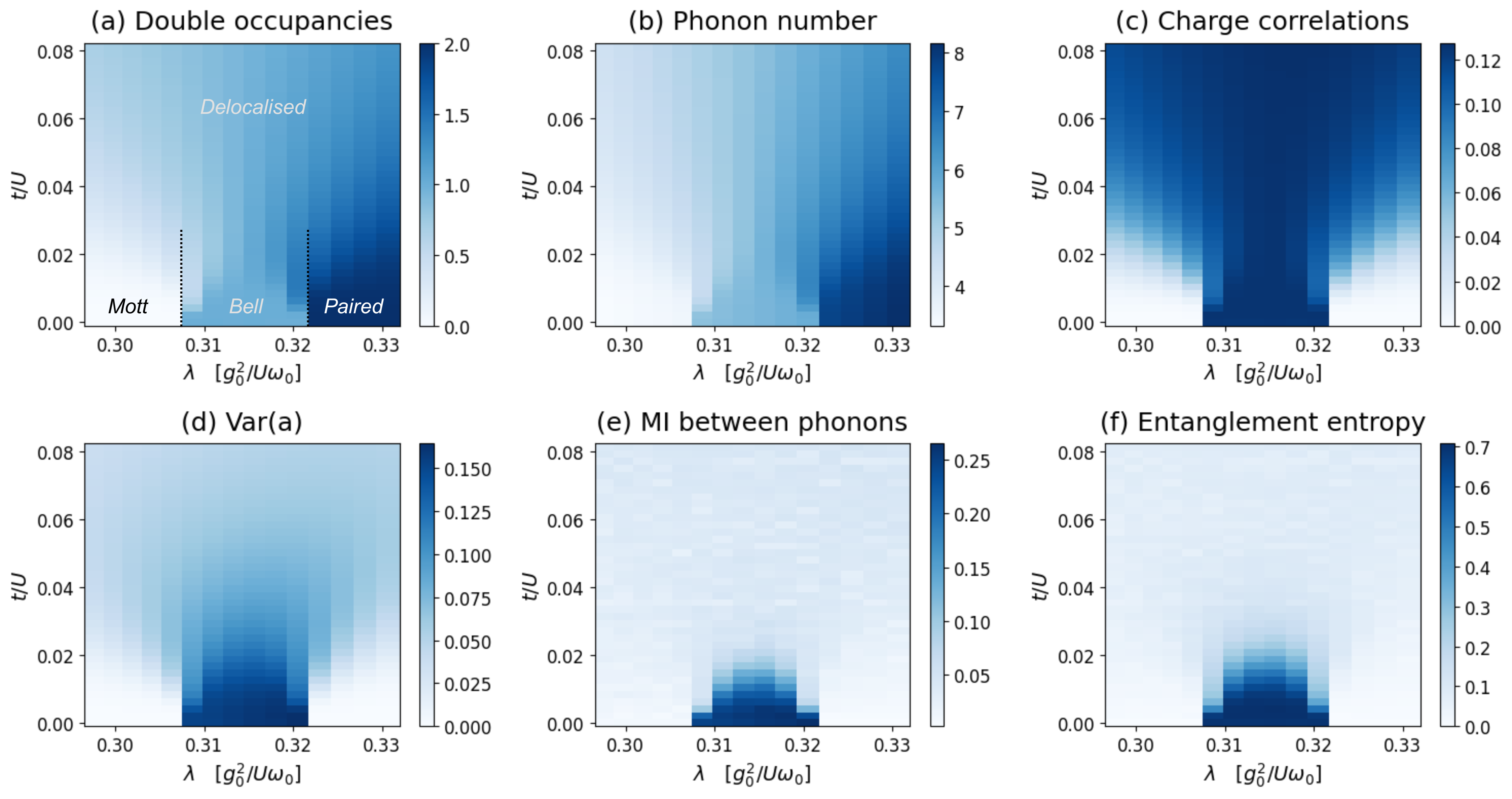}
    \caption{\textit{The phase spectrum of the two-tube system with intertube coupling strength $V/U=0.02$.} The figures correspond to the following observables: (a) the electronic double occupancies in a single tube, (b) the phonon number in a single tube, (c) the average electronic charge correlations in a single tube, (d) the variance of the phononic annihilation operator in a single tube, (e) the mutual information between phonons of different tubes, and (f) the entanglement entropy between subsystem $\mathcal A$ and $\mathcal B$. The entropic quantities were computed with a logarithm of base e.}
    \label{fig:heatmaps}
\end{figure*}

\begin{align}
    \hat H^{\scriptscriptstyle \text{LF}}&=-t\sum_{i}(\hat c_{i}^{\dagger}\hat c_{i+1}e^{\sum_{\mu}\frac{g_{i,\mu}-g_{i+1,\mu}}{\omega_{\mu}}(\hat a_{\mu}^\dagger - \hat a_{\mu})} + h.c.)\\
    &+ U\sum_{i}\hat n_{i,\sigma}\hat n_{i,\sigma'}+ \sum_{\mu} {\omega_{\mu}}{\hat a_{\mu}^\dagger}{\hat a_{\mu}} \underbrace{- \sum_{i,j,\mu}\frac{g_{i,\mu}g_{j,\mu}}{\omega_\mu}\hat n_i \hat n_j}_{\hat H_{\tilde{U}}}\nonumber
    \label{LF}
\end{align}

\noindent The LF transformation effectively replaces the electron-phonon interaction $\hat H_{e-p}$ by an attractive long-range electron-electron interaction $\hat H_{\tilde{U}}$ plus 
% written below:
% \begin{equation}
%     \hat{H}_{\tilde{U}} = - \sum_{\mu} \sum_{i,j} \frac{g_{i,\mu} g_{j,\mu}}{\omega_{\mu}} \hat{n}_i \hat{n}_j.
% \end{equation}
a phonon-mediated tunneling term. As previously noted, the attractive electron-electron interaction can stabilize the otherwise repulsive Coulomb interaction $\hat{H}_U$. Significant insight into the physics of the problem can be gained by discussing the expected physics in the atomic limit, where no hopping occurs. \\

\noindent\textit{Zero-tunneling regime--} 
At \( t = 0 \), the electronic and phononic degrees of freedom %
are effectively uncoupled, and the ground state (GS) of the system is determined by the interplay between the electronic components of the Hamiltonian: \( \hat{H}_U \), \( \hat{H}_{\tilde{U}} \), and $ \hat{H}_V$. 
Therefore, the phononic part of the GS of the system can be taken as the vacuum state. 
For the uncoupled tubes at
\(V = 0\), the only relevant parameter of the system is given by $\lambda = g_0^2/(U\omega_0)$, and the system trivially reduces to the single-CNT problem discussed in \cite{bhattacharya_phonon-induced_2021}, where two distinct GS configurations arise. In the weak-coupling regime (small values of \(\lambda\)), the  GS in each tube corresponds to a \emph{\enquote{Mott insulating state}} (\emph{M}) in which each QD hosts at most one electron. As the coupling strength increases, a transition occurs to a \emph{\enquote{Paired state}} (\emph{P}),  
where electrons pair in the inner QDs due to the attractive interaction mediated by the phonons. At $\lambda=\lambda_c$, a transition between both regimes occurs. Using Eq. (3), the critical coupling strength is $\lambda_c=3/\pi^2$ if infinitely many phononic modes are considered or  $\lambda_c=\pi^2/32$ if a single mode is considered.
For the uncoupled tube case, the GS of the system is simply a product state: 
\mbox{$\ket{GS}_{\scriptscriptstyle \mathcal{AB}} = \ket{M}_{\scriptscriptstyle \mathcal{A}} \ket{M}_{\scriptscriptstyle \mathcal{B}}$} in the weak-coupling regime; or \mbox{$\ket{GS}_{\scriptscriptstyle \mathcal{AB}} = \ket{P}_{\scriptscriptstyle \mathcal{A}} \ket{P}_{\scriptscriptstyle \mathcal{B}}$} in the strongly interacting limit.

As the inter-tube interaction is turned on, i.e., $V \neq 0$, the spectrum of the system changes significantly. Direct diagonalisation shows a new GS configuration emerging between the Mott and the Paired state, which breaks the symmetry within each tube and correlates the electronic states of the two nanotubes. It corresponds to a superposition of the (four) lowest energy degenerate electronic states: 
\begin{eqnarray}
\ket{GS}_{\scriptscriptstyle \mathcal{AB}} & = \frac{1}{2} (\ket{M_{\scriptscriptstyle \mathcal{A}}, P_{\scriptscriptstyle \mathcal{B}}} + \ket{P_{\scriptscriptstyle \mathcal{A}}, M_{\scriptscriptstyle \mathcal{B}}} \\ \nonumber
               & +\ket{I^l_{\scriptscriptstyle \mathcal{A}}, I^r_{\scriptscriptstyle \mathcal{B}}} + \ket{I^r_{\scriptscriptstyle \mathcal{A}}, I^l_{\scriptscriptstyle \mathcal{B}}} ).
\label{NGS1}
\end{eqnarray}

\noindent In the Fock occupation basis, the above states correspond to 
$ \ket{M}\equiv \ket{1,1,1,1};\ \ket{P}\equiv \ket{0,2,2,0};\ \ket{ I^l}\equiv \ket{1,2,1,0}$ and $\ket{I^r}\equiv \ket{0,1,2,1}$ (see Fig. \ref{fig:schematic}). The structure of $\ket{GS}_{\scriptscriptstyle \mathcal{AB}}$ is now a highly entangled electronic state between all lowest energy configurations. 
As previously mentioned, when the coupling strength is sufficiently high, phonons on each tube induce an effective attractive electron-electron interaction that favours the formation of electron pairs. However, this configuration is energetically penalised by the inter-tube Coulomb repulsion and the on-site potential. Stability is achieved when electrons are in a superposition of states with a single double occupation on average (see Eq.(4)). Since the (unitary) LF transformation commutes with the Coulomb interaction, this superposition state corresponds to the true GS of the system at \(t = 0\), where effectively phonons mediate an attractive electron-electron interaction. \\

\noindent\textit{Finite tunneling regime --}  
For $t\neq 0$, the GS of the system must be obtained numerically. However, this is challenging given the large size of the phononic Hilbert space. To address this issue, one can restrict the analysis to the so-called \emph{single-mode approximation}, where instead of the infinite sum of mode frequencies $\omega_\mu$, only the lowest mode ($\mu =1$) is considered. This approach has already been validated in the single-CNT case \cite{bhattacharya_phonon-induced_2021}.

Despite this simplification, the single-mode approximation remains computationally demanding. In the strong coupling regime, the number of phonons involved is large enough to render the associated Hilbert space size intractable. An estimate for realistic parameters at $t=0$ can be obtained using the bosonic creation and annihilation operators in the LF representation. The transformed annihilation operator is given by
$\hat{a}_{\mu}^{\scriptscriptstyle \text{LF}}=e^{\hat{S}}\hat{a}_{\mu}e^{-\hat{S}}
=\hat{a}_{\mu}-\sum_i (g_{i,\mu}/\omega_{\mu})\hat{n}_i$, which yields
\begin{equation}
    \langle N \rangle
    =
    \langle \hat{a}^{\dagger \scriptscriptstyle \text{LF}}\hat{a}^{\scriptscriptstyle \text{LF}} \rangle_{\scriptscriptstyle \text{GS}}
    \propto
    \left[
        \sum_i n_i \left(\frac{g_0}{\omega_0}\right)
    \right]^2
    \simeq 10^6 .
\end{equation}
Here, we have omitted the mode index $\mu$, since we focus only on the lowest mode. To reduce the number of phononic states and thus the size of the Hilbert space, an \emph{iterative shift method} can be used~\cite{bhattacharya_phonon-induced_2021}. This technique effectively corresponds to a displacement of the phononic state to a coherent state with a mean phonon number identical to that of the original state. The process is repeated iteratively until convergence is reached, allowing for a substantial truncation of the phononic subspace. 
However, even this shift method becomes ineffective in our case, as the presence of a second tube dramatically increases the phonon number at finite tunneling.

To address this new challenge, we adopt a novel strategy. Recalling that at $t=V=0$, the only relevant physical scale is $\lambda = g_0^2 / (U \omega_0) $,  we perform a rescaling of both the phonon frequency and the on-site Coulomb interaction, namely $\omega_0 \to \omega_0'$ and $U \to U'$, such that the product   $\omega_0' U' = \omega_0 U$ remains constant. This transformation effectively reduces the phonon number and thereby the number of relevant phononic states, rendering the problem numerically tractable while preserving the essential features of the original model. For the rescaling, we chose  \( \omega_0' = 10^3 \omega_0 \) and \( U' = 10^{-3} U \) which leads to a significantly reduced phonon number of \( N  \simeq 10 \).

To validate our approach, we compare the results obtained using our rescaling method with those from the shift method proposed in~\cite{bhattacharya_phonon-induced_2021} for the single-tube case. All expected features are correctly reproduced at the appropriate values of the Hamiltonian parameters, although our method consistently shifts transitions to lower tunneling values, as can be expected from a closer inspection of Eq.(3). With this novel technique, we are now able to perform a full numerical analysis of the system while truncating the phononic subspace to approximately 50 states. All details of the numerical methods are provided in the Supplementary Material (SM).

Notably, adopting the single-mode approximation ($\mu = 1$) in our analysis lifts the degeneracy between the Mott-Paired states and the intermediate states (see SM). The GS in the novel phase reduces to:
\begin{align}
    \small
% \ket{GS}_{AB} = \frac{1}{2}\left(\ket{M}_{\scriptscriptstyle A}\ket{P}_{\scriptscriptstyle B}\otimes\ket {\mu=1,N_{\scriptscriptstyle M}}_{\scriptscriptstyle A}\ket{\mu=1,N_P}_{\scriptscriptstyle B}\right)
\ket{GS}_{\mathcal{AB}} &= \frac{1}{2}\Bigl(\ket{M}\ket {N_{\scriptscriptstyle M}}_{ \mathcal{A}}\otimes\ket{P}\ket{N_{\scriptscriptstyle P}}_{ \mathcal{B}}\\
&\quad  + \ket{P}\ket{N_{\scriptscriptstyle P}}_{\mathcal{A}} \otimes \ket{M}\ket {\small{ N_{\scriptscriptstyle M}}}_{\mathcal{B}} \Bigr)\nonumber
\label{NGS2}
\end{align}

\noindent where $\ket{N_{\scriptscriptstyle M/P}}$ denotes the phononic state with an occupation number associated with the electronic configurations $M/P$ in systems $\mathcal{A}/\mathcal{B}$, respectively. Note that these phononic states are not fully orthogonal.
%we have now included the phononic part where the occupation number, N, of the fundamental mode is dependent on the electronic configuration. 
For greater readability, we separate the subsystems by a tensor product symbol. We refer to this configuration as an electronic Bell state. 

\begin{figure*}[t!]
    \centering
    \includegraphics[width=\linewidth]{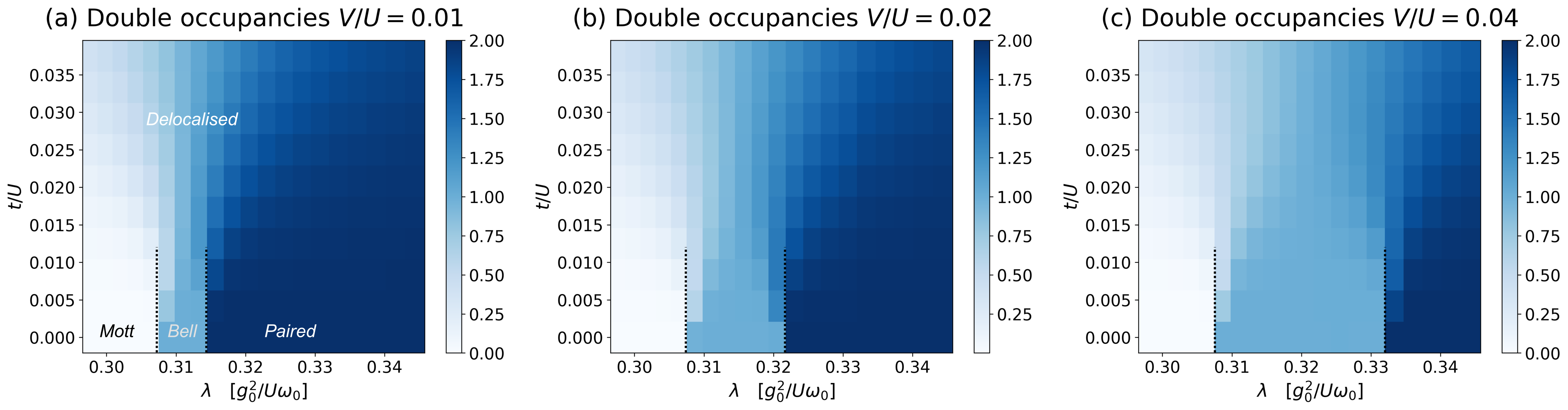}
    \caption{\textit{Phase diagram dependence on the inter-tube coupling $V/U$}. We display the number of electronic double occupations as a function of the inter-tube coupling:  (a1) $V/U = 0.01$, (a2) $V/U = 0.02$, and  (a3) $V/U=0.04$. The transition from the Mott to Bell phase occurs at the same critical value regardless of the value of $V/U$, while the transition from Bell to Paired phase shifts to larger values of $\lambda$ as $V/U$ increases. Notably, the Bell phase, characterized by an average double occupancy of one, expands linearly with increasing inter-tube coupling strength.}
    \label{fig:increasedcoupling}
\end{figure*}

In Fig. \ref{fig:heatmaps}, we present our numerical results for the two-tube model at finite tunneling \( t \) and inter-tube Coulomb coupling \( V/U = 0.02 \). The observables span electronic, phononic, and coupled degrees of freedom of the numerically obtained ground state configurations. For sufficiently small values of \( t/U \), and in agreement with the LF transformation, three distinct GS configurations can be identified via the electronic double occupancies, as shown in Fig. \ref{fig:heatmaps}(\textit{a}). We identify them as Mott (M), Bell (B), and Paired (P) phases, characterized by average electronic double occupancies of zero, one, and two, respectively. 
A similar parameter dependence is obtained from the phonon number occupation: $N = \langle \hat{a}^{\dagger} \hat{a} \rangle_{\scriptscriptstyle\text{GS}}$, as shown in Fig. \ref{fig:heatmaps}(b). The Mott phase corresponds to low phonon occupation, so the repulsive on-site Coulomb interaction dominates. The Paired phase, on the other hand, exhibits a high number of energetic phonons (associated with the energy scale \( \hbar \omega_0 \)) and electron-phonon interactions induce attractive electron-electron interactions. The mean phonon number in the Bell phase corresponds to the average between the two above extremes.  As tunneling increases, both the double occupancy and phonon number change gradually along the full range of the parameter.
%, which quantifies the ratio of electron-phonon interaction strength \( (g_0 / U \omega_0) \) to Coulomb repulsion \( U \). 
This behavior indicates the emergence of a delocalized (or superfluid-like) phase, where neither the average electron number on each QD nor the average phonon number is well defined. 

A better characterization of the delocalized phase (and the Bell phase) is given by analyzing the average electronic charge correlation within a single tube, $C=\sum_{i,j}\langle n_i n_j\rangle_{\scriptscriptstyle\text{GS}}-\langle n_i\rangle_{\scriptscriptstyle\text{GS}} \langle n_j\rangle_{\scriptscriptstyle\text{GS}}$. As expected, for the phases corresponding to a well-defined number of particles in a product state, the electronic correlation function is zero, as in the Mott and Paired phases, while it is maximal %when the average number is in a correlated state, as 
in the Bell or the delocalized phase.  A similar characterization can be obtained from analyzing the variance of the phononic annihilation operator $\text{Var}(\hat a)\equiv \langle\hat a^\dagger\hat a\rangle_{\scriptscriptstyle\text{GS}} - \langle\hat a^\dagger\rangle_{\scriptscriptstyle\text{GS}} \langle\hat a\rangle_{\scriptscriptstyle\text{GS}}$, a measure for the classical correlations of the phonons within each tube. However, the properties of the Bell phase differ remarkably from those of the delocalized phase. To illustrate this, we also choose the mutual information and the entanglement entropy as observables.

The quantum mutual information $I(\mathcal{A},\mathcal{B})$ is a measure of correlations between two systems $\mathcal{A}$ and $\mathcal{B}$, which includes both classical and quantum correlations \cite{schumacher_quantum_1996} and is defined as:
\begin{equation}
    I(\mathcal{A},\mathcal{B})= S(\mathcal{A})+S(\mathcal{B})-S(\mathcal{A},\mathcal{B})
\end{equation}
where  $S(\mathcal{A})=-\text{Tr} (\rho_{\scriptscriptstyle\mathcal{A}}\log\rho_{\scriptscriptstyle\mathcal{A}})$ is the von Neumann entropy of a system described by a density matrix $\rho_{\scriptscriptstyle\mathcal{A}}$, and $\rho_{\scriptscriptstyle\mathcal{A}}=\text{Tr}_{\scriptscriptstyle\mathcal{B}} (\rho_{\scriptscriptstyle \mathcal{AB}})$. In Figure \ref{fig:heatmaps}(e), we display the mutual information between phononic degrees of freedom: $\mathcal{I}(\mathcal{A},\mathcal{B})_{p} = \mathcal{S}(\mathcal{A})_p+S(\mathcal{B})_p-S(\mathcal{A},\mathcal{B})_{p}$ where $p$ denotes the phononic degrees of freedom. The two tubes are correlated in the Bell phase and only in this phase. Once we have discarded the electronic degrees of freedom, one way to infer if phononic correlations are classical or quantum is to compute the negativity in the state $\rho_{p}$, which is a measure of entanglement in bipartite splittings of arbitrary dimensions \cite{vidal_computable_2002}. The negativity is zero for the whole range of parameters, including the ones defining the Bell phase. This is expected, since within the single-mode approximation the only phononic degree of freedom is the occupation number N of the fundamental mode. This indicates that the two carbon nanotubes A and B are classically correlated due to the intertube Coulomb repulsion. Finally, we complete this analysis by calculating the entanglement entropy of the electronic degrees of freedom, $E(\ket{GS}_{\scriptscriptstyle\mathcal{AB}})=-\text{Tr}(\rho_{\scriptscriptstyle\mathcal{A}} \log\rho_{\scriptscriptstyle\mathcal{A}})$, which is nothing other than the von Neumann entropy of one of the subsystems. As expected, we find that the electronic degrees of freedom are strongly quantum correlated in the Bell phase.

We conclude our study by analyzing the effect of the inter-tube Coulomb coupling $V$  in the phase diagram as shown in  Figure \ref{fig:increasedcoupling}. Notably, the transition from the Mott to the Bell regime is not affected by the value of $V$ and always occurs at the same critical value $\lambda_c$. This is because both the Mott states and Bell states experience the same energy shift of $4V/U$. Increasing the value of $V$ does, however, affect the transition between the Bell and Paired phases, extending the former over a larger range of $\lambda$'s, but all the properties displayed in Fig. \ref{fig:heatmaps} remain the same. We note that coupling strengths as large as $V/U\sim1$ are realistic, which would further enlarge the Bell phase; in the zero-tunneling limit within the single-mode approximation, this corresponds to the range $\lambda=0.308$ to $\lambda=0.925$.\\

\noindent \textit{{Summary --}} We have proposed a quasi 2D setup of an electro-nanomechanical quantum simulator using two suspended nanotubes hosting quantum dots. We have theoretically derived the phase diagram of the system as a function of the relevant coupling parameters and demonstrated the presence of novel maximally entangled (electronic) Bell states arising from the interplay between geometry (quasi 2D), mobility of electrons (tunneling), Coulomb repulsion, and strong electron-phonon interactions. Our theoretical findings are based on a meaningful truncation of the phononic Hilbert space, which allows us to analyse in detail entanglement and correlations between electronic, phononic and electron-phonon degrees of freedom. This analysis focuses on the core physics and omits device-level nonidealities—including alignment tolerances between CNTs and the possibility of nonidentical dot sizes. Systematic exploration of these factors is left for future work. However, we expect the phase diagram presented here to remain robust against moderate misalignment of the quantum dots. An interesting open question is what novel phenomena may emerge when additional QDs are added to each tube. In the limits of weak and strong electron–phonon coupling, the ground state of the system is known to correspond to a Mott-insulating state and an all-paired state, respectively. Between these two limits, additional phases are expected to arise, characterized by different numbers of double occupations and distinct correlation patterns. Future work should also examine whether the newly identified phase remains robust when the system is coupled to its environment. Another intriguing direction is to determine whether the proposed architecture could provide a viable platform for realizing phononic quantum correlations, although addressing this question would require going beyond the single-mode approximation. At present, both directions remain challenging because of numerical intractability. Single suspended CNT quantum dots have been successfully demonstrated with the required electromechanical coupling strengths $g_0/(2\pi\hbar)$ up to 800~MHz and mechanical frequencies in the tens-of-MHz range. 
Having controlled, beyond-one-dimensional strongly correlated electron–phonon systems at our disposal is fundamental for the understanding of electrical and mechanical properties of quantum materials. \\

\noindent \textit{Acknowledgments --}
We acknowledge stimulating discussions with M. Lewenstein, U. Bhattacharya, L. Zhang, and J. Niwemuto.  DU and AS acknowledge financial support from MICINN grant PID2022-139099NBI00, with the support of FEDER funds, the Spanish Government with funding from European Union NextGenerationEU (PRTR-C17.I1), the Generalitat de Catalunya, the Ministry for Digital Transformation and of Civil Service of the Spanish Government through the QUANTUM ENIA project -Quantum Spain Project- through the Recovery, Transformation and Resilience Plan NextGeneration EU within the framework of the Digital Spain 2026 Agenda. MC, SF, and AB acknowledge support from ERC Advanced Grant 101198268-QTube, Marie Sklodowska-Curie grant agreement No. 847517 and 101105814, MICINN Grant No. RTI2018-097953-B-I00 and PID2021-122813OB-I00, the Quantera grant (PCI2022-132951), the Fondo Europeo de Desarrollo, the Spanish Ministry of Economy and Competitiveness through Quantum CCAA, TED2021-129654B-I00, EUR2022-134050, and CEX2019-000910-S [MCIN/AEI/10.13039/501100011033], MCIN with funding from European Union NextGenerationEU(PRTR-C17.I1), Fundacio Cellex, Fundacio Mir-Puig, Generalitat de Catalunya through CERCA, and 2021 SGR 01441.

\section{Supplementary Material}
\label{appendix}
\noindent \emph{Single mode approximation --} 
In the single-tube setup, the single-mode model proves to be a valuable approximation. The overall phase spectrum is not affected by the exclusion of higher modes, although accounting for these modes is computationally demanding. For example, in the zero-tunneling limit, both models exhibit a single phase transition from a Mott insulating state to the Paired state, with only a slight shift in the critical coupling value $\lambda_c$. As discussed previously, applying the single-mode model to the two-tube system lifts the degeneracy between the intermediate states and the Mott–Paired state within the Bell phase. Nevertheless, just like when all modes are included, two phase transitions are observed, with the Mott and Paired states serving as the GSs in the weak- and strong-coupling limits, respectively. Moreover, an additional phase emerges in between, characterized by strong entanglement between the two subsystems. We therefore conclude that the single-mode model remains a valuable approximation for the two-tube setup.\\

\noindent \emph{Iterative Shift method --} 
We previously mentioned that we adapted the iterative shift method introduced in \cite{bhattacharya_phonon-induced_2021} to the two-dimensional system. The physical interpretation behind this transformation is the reduction of the fluctuations in the phononic space by mapping to a coherent state. For $t\neq 0$, one iteratively updates the shift parameter to meet the following two requirements: 1) The state must have the same expected number of phonons as the original state; 2) The state must be the GS of the transformed system. While the transformed system is different, for many relevant observables, e.g. the number of electronic double occupancies or the average charge correlations, the expected number of phonons is the relevant parameter alongside to the e-p coupling strength. Yet the number of contributing phononic states is minimal, and the method allows for an efficient truncation of the phononic subspace.\\

\begin{figure}
    \centering 
    \includegraphics[width=\linewidth]{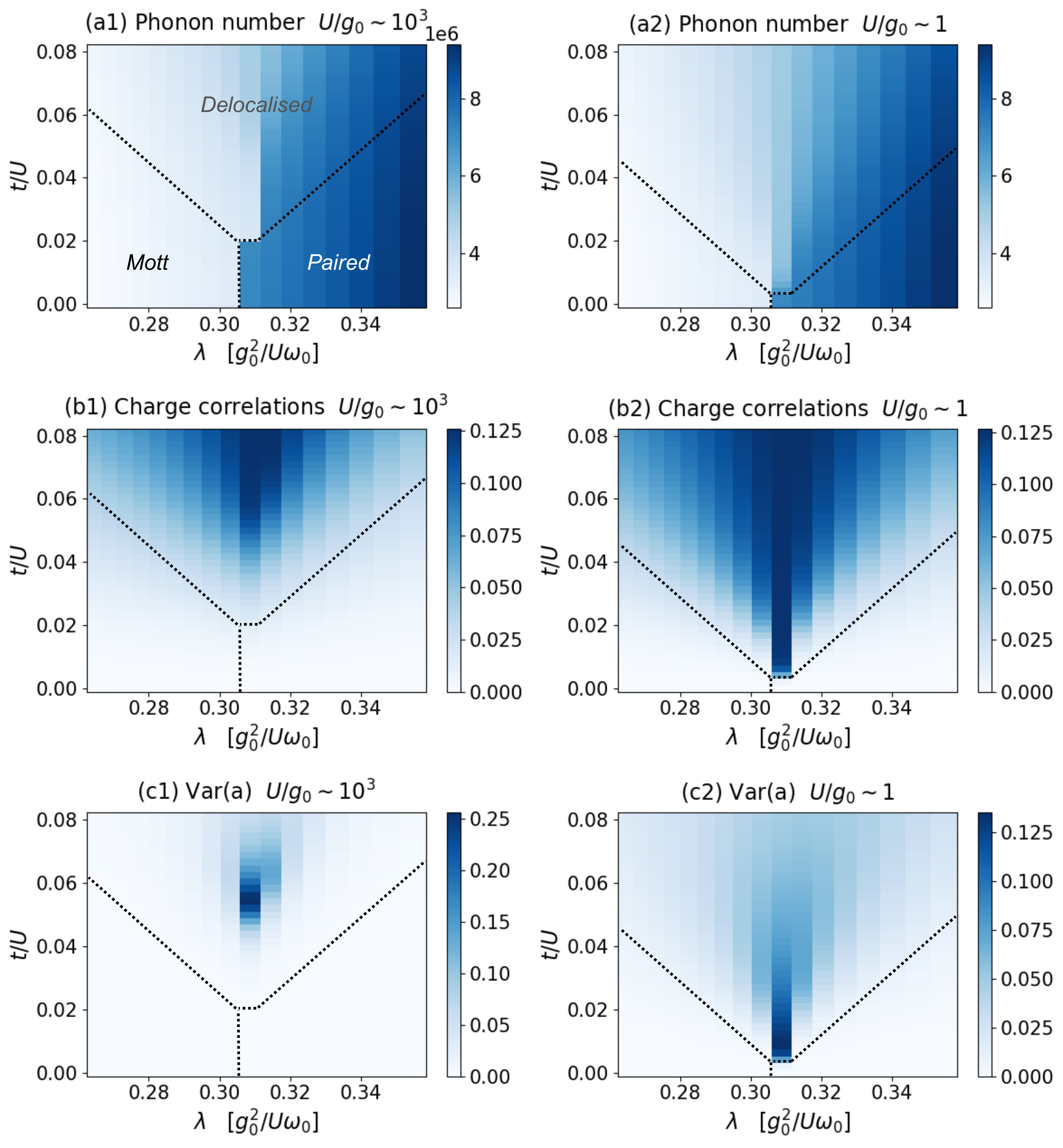}
    \caption{\textit{The effect of the rescaling}. The phase spectrum of the one-dimensional setup is displayed for a realistic system (\textit{left}) and a system of low phonon numbers (\textit{right}), exemplified on the phonon number (\textit{top}), the average charge correlations (\textit{middle}) and the variance of the phononic annihilation operator (\textit{bottom}).}
    \label{fig:hypothetical}
\end{figure}

\noindent \emph{Rescaling of frequency and on-site potential --} The rescaling of these parameters allows for an efficient computation of the GS by effectively reducing the phonon number present. While such systems differ from those with experimentally realistic parameters, they share many key features with the latter and are therefore a meaningful approximation. Fig. \ref{fig:hypothetical} shows the phase spectrum of the one-dimensional setup for a realistic system (\textit{left}) and a low-phonon-number system (\textit{right}) illustrated on the phonon number (\textit{top}), the average charge correlations (\textit{middle}) and the variance of the phononic annihilation operator (\textit{bottom}). We notice that in the zero-tunneling limit, the two occurring phases, as well as the phase transition from Mott to Paired, are unaffected, apart from the absolute number of phonons present. 
Further, the behavior along the $\lambda$-axis is preserved. Two differences between the realistic system and the one with rescaled parameters are the total number of phonons and the absolute values of the variance of the phononic operator, yet the relative difference is well comparable. However, the main effect, as indicated by the electronic and phononic correlations, is that the delocalized regime is reached for smaller values of $t/U$ for the rescaled system. 
This vertical distortion of the spectrum can be explained by taking a closer look at the tunneling term of the Lang-Firsov Hamiltonian
\begin{align*}
    \hat H^\mathcal{L}_t=-t\sum_{i}\hat c_{i}^{\dagger}\hat c_{i+1}e^{\sum_{\mu}\frac{g_{i,\mu}-g_{i+1,\mu}}{\omega_{\mu}}(\hat a_{\mu}^\dagger - \hat a_{\mu})} + h.c. 
\end{align*}
As previously discussed, this term acquires an additional phonon-dependent phase that is affected by the rescaling of the frequency $\omega_\mu$.
Overall, this allows the previously displayed results for the two-tube system to be mapped to more realistic systems. Notably, in these systems, the Bell phase expands further, with strong correlations between the subsystems persisting even at higher tunneling amplitudes.

\bibliography{bib}
\end{document}